\documentclass[preprint,prl,showpacs,preprintnumbers,superscriptaddress,amsmath,amssymb,letter]{revtex4-1}

\usepackage{units}
\usepackage{graphicx}

\begin{document}

\title{Magnetic Moments of Chromium-Doped Gold Clusters: \\ The Anderson Impurity Model in Finite Systems}

\author{K.~Hirsch}
\email{konstantin.hirsch@physik.tu-berlin.de}

\author{V.~Zamudio-Bayer}
\author{A.~Langenberg}
\author{M.~Niemeyer}
\author{B.~Langbehn}

\affiliation{Institut f\"ur Optik und Atomare Physik, Technische Universit\"at Berlin, Hardenbergstra{\ss}e 36, 10623 Berlin, Germany}
\affiliation{Institut f\"ur Methoden und Instrumentierung der Forschung mit Synchrotronstrahlung, Helmholtz-Zentrum Berlin f\"ur Materialien und Energie GmbH, Albert-Einstein-Stra{\ss}e 15, 12489 Berlin, Germany}

\author{T.~M\"oller}
\affiliation{Institut f\"ur Optik und Atomare Physik, Technische Universit\"at Berlin, Hardenbergstra{\ss}e 36, 10623 Berlin, Germany}

\author{A.~Terasaki}
\affiliation{Cluster Research Laboratory, Toyota Technological Institute, 717-86 Futamata, Ichikawa, Chiba 272-0001, Japan}
\affiliation{Department of Chemistry, Kyushu University, 6-10-1 Hakozaki, Higashi-ku, Fukuoka 812-8581, Japan}

\author{B.~v.~Issendorff}
\affiliation{Fakult\"at f\"ur Physik, Universit\"at Freiburg, Stefan-Meier-Stra{\ss}e 21, 79104 Freiburg, Germany}

\author{J.~T.~Lau}
\email{tobias.lau@helmholtz-berlin.de}
\affiliation{Institut f\"ur Methoden und Instrumentierung der Forschung mit Synchrotronstrahlung, Helmholtz-Zentrum Berlin f\"ur Materialien und Energie GmbH, Albert-Einstein-Stra{\ss}e 15, 12489 Berlin, Germany}

\begin{abstract}
The magnetic moment of a single impurity atom in a finite free electron gas is studied in a combined x-ray magnetic circular dichroism spectroscopy and density functional theory study of size-selected free chromium-doped gold clusters. The observed size-dependence of the local magnetic moment can essentially be understood in terms of the Anderson impurity model. Electronic shell closure in the host metal minimizes the interaction of localized impurity states with the confined free electron gas and preserves the full magnetic moment of $\unit[5]{\mu_B}$ in $\mathrm{CrAu}_{2}^{+}$ and $\mathrm{CrAu}_{6}^{+}$ clusters. Even for open-shell species, large local moments are observed that scale with the energy gap of the gold cluster. This indicates that an energy gap in the free electron gas generally stabilizes the local magnetic moment of the impurity.
\end{abstract}


\preprint{}

\date{\today}

\maketitle

\begin{figure}[t]
\includegraphics{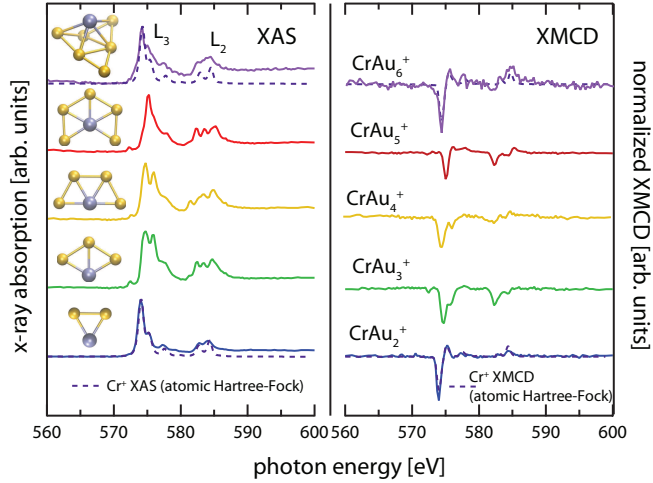}
\caption{\label{fig:CrAu_XMCD}Left: Linear x-ray absorption spectra of $\mathrm{CrAu}_{n}^+$ clusters, normalized to the $L_3$ maximum intensity.  Right: XMCD spectra of $\mathrm{CrAu}_{n}^{+}$, $\mathrm{n=2-6}$, normalized to the number of unoccupied $3d$ states. In both cases, the spectra of $\mathrm{CrAu}_{2}^+$ and $\mathrm{CrAu}_{6}^+$ are overlaid with theoretical spectra of isolated $\mathrm{Cr^+}$ (dashed line) from a Hartree-Fock calculation. Also shown are relaxed ground state structures \cite{Torres05a} (dark atom: chromium; light atoms: gold).\@}
\end{figure}

The interaction of localized impurity states with a free electron gas \cite{Anderson61} leads to complex phenomena such as Friedel oscillations \cite{Friedel58} or the Kondo effect \cite{Kondo64}.\@ The properties of magnetic impurities in non-magnetic bulk metals \cite{deHaas34} have therefore been subject of intense research over the last 50 years.
Considerable advances in the understanding of these many-body effects have been made by applying photo\-emission, x-ray magnetic circular dichroism (XMCD), and scanning tunneling spectroscopy to adatoms \cite{Madhavan98,Li98c,Gambardella02,Heinrich04,Wahl04,Carbone10}, clusters on surfaces \cite{Neel08}, or well defined quantum dots \cite{Cronenwett98,GoldhaberGordon98}.\@ This allows the study of basic parameters such as the on-site Coulomb repulsion or the amount of interaction of the impurity atom with the host metal. In all these cases, however, the impurity atom is in contact with a bulk free electron gas that has a continuous density of states. 
In contrast, the study of single impurities in size-selected clusters would allow to characterize the interaction of localized electronic states with a finite electron gas, i.e., with a well defined number of electrons occupying a highly discrete density of states.  
Because this introduces a new parameter for the control of electronic and magnetic properties, isolated doped coinage-metal clusters have been studied intensively by density functional theory (DFT) calculations \cite{Torres05a,Yuan05} and experiment: Their electronic and geometrical structure as well as their relative stability have been probed by photoelectron spectroscopy \cite{Koyasu02,Li05b,Tono05}, infrared dissociation \cite{Lin10b,Lin10a} and ultraviolet photofragmentation \cite{Neukermans03,Janssens05a}, as well as electron diffraction \cite{Wang09}.\@ Yet, none of these experimental techniques directly addresses the magnetic properties. 
Here, we use chromium-doped gold clusters as model systems that combine considerable local magnetic moments, carried by the $3d$ electrons of the impurity atom, with a finite free electron gas formed by the gold host. We study the magnetic moment of a single chromium impurity by local and element-specific XMCD spectroscopy of size-selected gas phase clusters \cite{Peredkov11b,Niemeyer12}.\@  
The results of our combined experimental and theoretical study show that the size-dependent change of the local magnetic moment in the finite, isolated system $\mathrm{CrAu}_{n}^{+}$ is correlated with the energy gap at the Fermi energy of the host cluster and can essentially be understood within the Anderson impurity model \cite{Anderson61}.\@
\newline
Experimentally, x-ray absorption and XMCD spectra of free size-selected doped gold clusters were taken in ion yield mode with a combined linear ion trap and superconducting solenoid setup \cite{Terasaki07,Hirsch09,Niemeyer12} with sufficient sensitivity to study singly doped size-selected clusters \cite{Lau09a,Lau11}.\@ The cluster beam is produced in a magnetron gas aggregation source by co-sputtering of chromium and gold targets, mass selected in a quadrupole mass filter, and transferred into the radio frequency ion trap by electrostatic and radio-frequency ion guides. Inside the ion trap, the doped gold clusters are magnetized by an external \unit[5]{T} magnetic field under continuous helium buffer gas cooling to an ion temperature of $\unit[\left(20\pm 5 \right)]{K}$.\@ The density of the dilute gas phase sample is $\unit[\approx 5\times 10^7]{ions \, cm^{-3}}$.\@ Resonant photoexcitation at the $L_{2,3}$ absorption edges of chromium was performed at BESSY II beamline UE52-SGM.\@ After x-ray absorption, the core-excited cluster ions relax via cascading Auger decay, which leads to photofragmentation of the parent ion. These photoion yield curves were taken for linear as well as for circular polarization with parallel and antiparallel alignment of photon helicity and magnetic field. The resulting x-ray absorption and XMCD spectra, shown in Fig.\ \ref{fig:CrAu_XMCD}, were normalized to the incident photon flux and scaled to unity at the $L_3$ edge for comparison. In addition, the XMCD spectra were normalized to the number of unoccupied $3d$ states as inferred from the integrated x-ray absorption signal. These x-ray absorption and XMCD spectra allow us to obtain local and element specific information on the electronic and magnetic structure of the chromium impurity in $\mathrm{CrAu}_{n}^{+}$ clusters. This holds true even though the XMCD spin sum rule \cite{Carra93} is not applicable to chromium because of the intermixing of $2p_{3/2} \rightarrow 3d$ and $2p_{1/2} \rightarrow 3d$ transitions. Nevertheless, quantitative information on the local magnetic moments can be obtained from a fingerprint analysis by comparison of calculated and experimental signatures in x-ray absorption and XMCD spectra. 
\newline
To analyze the electronic and magnetic properties of $\mathrm{CrAu}_{n}^{+}$, established ground state geometries \cite{Torres05a} were re-optimized in a Kohn-Sham DFT framework as implemented in the \textsc{quantum espresso 5.0} plane wave code \cite{Giannozzi09}, employing the Perdew-Burke-Ernzerhof approximation to the exchange-correlation functional \cite{Perdew96}.\@ The kinetic energy cut-off for wave functions and charge density were set to \unit[680]{eV} and \unit[2720]{eV}, respectively. Scalar relativistic effects were taken into account by using an ultra-soft pseudopotential of the Vanderbilt type \cite{Vanderbilt90}.\@ In addition, atomic Hartree-Fock calculations of x-ray absorption and XMCD spectra for an isolated $\mathrm{Cr^+}$ ion with $\mathrm{[Ar]}\,3d^5$ initial and $2p^5\,3d^6$ final state configurations were performed in the \textsc{missing} program package \cite{Cowan81}.\@
\begin{figure}[t]
\includegraphics{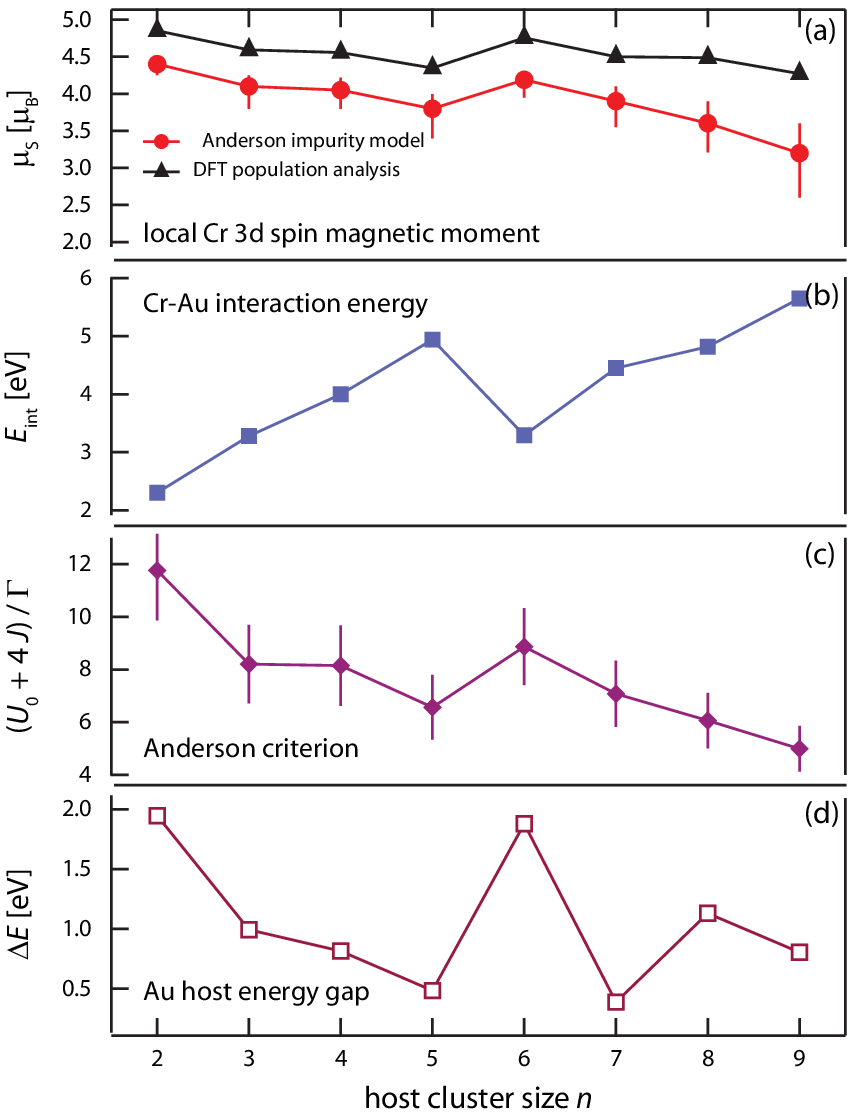}
\caption{\label{fig:onsiteU} (a) Calculated local chromium spin magnetic moments $\mu_S$ of $\mathrm{CrAu}_{n}^{+}$ from a DFT population analysis (black triangles) and as predicted in the Anderson model (red bullets); (b) Cr-Au interaction energy $E_{\mathrm{int}}$; (c) Anderson criterion $\left( U_{0} + 4 J \right ) / \Gamma$; (d) energy gap $\Delta E$ of the gold host.}
\end{figure}
\newline
As can be seen in Fig.\ \ref{fig:CrAu_XMCD}, all $\mathrm{CrAu}_{n}^{+}$ clusters under study exhibit local and total magnetic moments, which are aligned in the presence of the external magnetic field, and therefore lead to nonvanishing XMCD signals independent of the detailed geometric and electronic structure. Furthermore, there is very good agreement of the calculated atomic $\mathrm{Cr^+}$ x-ray absorption spectrum \cite{Hirsch12a,Lau09b} with the experimental spectra of $\mathrm{CrAu_2^+}$ and, to a lesser extent, $\mathrm{CrAu_6^+}$, indicating atomic-like chromium $3d^5$ configurations in both cases. This strong localization of the $3d$ electrons at the chromium site is also reflected in screening that shifts the $\mathrm{CrAu_{2,6}^+}$ $L_3$ absorption line by $\unit[\approx 0.5]{eV}$ to lower excitation energy \cite{Lau08,Hirsch12b} as compared to $\mathrm{CrAu_{3-5}^+}$.\@ The XMCD spectra of $\mathrm{CrAu}_{n}^{+}$ corroborate this interpretation: Again, the calculated XMCD spectrum of ionic $\mathrm{Cr^+}$ shows excellent agreement with that of $\mathrm{CrAu_2^+}$, and good agreement with the XMCD spectrum of $\mathrm{CrAu_6^+}$ within the experimental signal-to-noise ratio. 
For $\mathrm{CrAu_2^+}$, a fit of the calculated XMCD asymmetry to the experimental spectrum yields an alignment of 0.42, which corresponds to the Brillouin value for a total spin $S = 5/2$ at the experimental conditions of $\unit[B = 5]{T}$ and $\unit[ T = \left(20 \pm 5 \right)]{K}$.\@ 
From this agreement, a localized, atomic-like $3d^5$ electron configuration and thus a local spin magnetic moment of $\unit[5]{\mu_B}$ can be deduced for the single chromium impurity in both clusters. This is also supported by DFT calculations as can be seen in Fig.\ \ref{fig:onsiteU} (a): In accordance with previous theoretical results \cite{Torres05a} we find local chromium spin magnetic moments of $\unit[4.85]{\mu_B}$ for $\mathrm{CrAu_{2}^+}$ and $\unit[4.75]{\mu_B}$ for $\mathrm{CrAu_{6}^+}$ by projecting the Kohn-Sham orbitals onto atomic wave functions. 
\newline
The situation is different for the experimental spectra of $\mathrm{CrAu}_{3-5}^{+}$ in Fig.\ \ref{fig:CrAu_XMCD}, where no agreement with atomic Hartree-Fock calculations can be found. This suggests that the chromium $3d$ electrons hybridize with gold $5d/6s$ states. Apparently, this hybridization leads to a reduction but not to a complete quenching of the local spin moment. 
\newline
This particular behavior of $\mathrm{CrAu_{2}^+}$ and $\mathrm{CrAu_{6}^+}$ can be understood from the size-dependence of the chromium-gold interaction. The gold subunits of both clusters, depicted in Fig.\ \ref{fig:CrAu_XMCD}, are structurally close to pure $\mathrm{Au_2}$ and $\mathrm{Au_6}$ \cite{Haekkinen00}, i.e., they remain nearly undistorted when adding the chromium impurity. In $\mathrm{CrAu_{3-5}^+}$, in contrast, the number of Cr-Au bonds is maximized, and the gold host is strained and deformed in comparison to its isolated, relaxed counterpart. This enhanced stability of $\mathrm{Au}_{2}$ and $\mathrm{Au}_{6}$ stems from shell closure for two and six delocalized $6s$ electrons \cite{Haekkinen00}, which, in spite of a strong $spd$ hybridization \cite{Haekkinen08}, form a free electron gas confined in a two dimensional potential well \cite{Janssens03}.\@ Therefore, $\mathrm{Au_{2,6}}$ are known to feature large second differences in binding energy and wide energy gaps $\Delta E$ of $\unit[\approx 2]{eV}$ at the Fermi energy \cite{Haekkinen00}.\@ 
Consequently, the chromium cation can be expected to interact more weakly in these two cases than with the open-shell gold clusters. This is indeed true, as can be inferred from the chromium-gold interaction energy $E_{\text{int}} = E \left( \mathrm{Cr^+}\right) + E \left( \mathrm{Au_n}\right) - E \left( \mathrm{CrAu}_{n}^{+}\right)$.\@ To extract the contribution of the chromium interaction with the gold cluster from the total energy, $E \left( \mathrm{Au}_{n}\right)$ is calculated in the same geometric configuration of $\mathrm{Au}_{n}$ as in $\mathrm{CrAu}_{n}^{+}$.\@ 
As expected, the weakest impurity-host interactions of \unit[2.5]{eV} and \unit[3.3]{eV} are indeed found for $\mathrm{CrAu_{2}^+}$ and $\mathrm{CrAu_{6}^+}$, respectively. In $\mathrm{CrAu_{3-5}^+}$, $E_{\mathrm{int}}$ increases to \unit[3.5--5]{eV} and indicates an increasing amount of covalent or metallic bonding, as can be seen in Fig.\ \ref{fig:onsiteU} (b).\@ In contrast to electronic shell closure, the coordination of the impurity atom only plays a minor role in $\mathrm{CrAu}_{n}^+$ even though it does lead to a slight decrease of the spin magnetic moment with increasing coordination number for two-dimensional clusters with $n = 2 - 5$, where the coordination number is equal to the number of gold atoms and the average chromium-gold distances of $\unit[\approx 2.7]{{\AA}}$ are comparable. A similar effect is observed for larger ($n = 6 - 9$) three-dimensional clusters. 
\newline
The interaction between the magnetic impurity and the free-electron host states is a crucial ingredient for the impurity magnetic moment in the Anderson impurity model \cite{Anderson61}.\@ Within this model, the size of the magnetic moment of the impurity atom sensitively depends on the interplay of the on-site Coulomb repulsion, i.e., the direct Coulomb interaction of two electrons in the same localized orbital, and on the width $2 \Gamma$ of the localized state. 
This width is determined by the coupling strength to the free electron states and by their density in the vicinity of the energy of the impurity state \cite{Anderson61}.\@ Since the presence of an energy gap $\Delta E$ strongly affects this density of states and thus the amount of hybridization, a relation between $\Delta E$ and the impurity magnetic moment can be anticipated. 
In the absence of interaction with the free electron gas, the impurity states $E$ and $E + U_{0}$ are separated by the bare Coulomb interaction $U_{0}$ that preserves the local magnetic moment if $U_{0}$ pushes the state $E + U_{0}$ above the Fermi level. In the presence of interaction, virtual states are formed at energies $E + U_{0}\cdot n_-$ and $E + U_{0}\cdot n_+$, where $n_{\pm}$ are the occupation numbers of the impurity atom majority and minority states. 
The separation of the virtual levels is reduced to an effective value $U_{\mathrm{eff}} = U_{0}\left(n_+ - n_-\right)$ by hybridization of the localized impurity states with free electron gas states, which causes a broadening of the virtual states \cite{Anderson61}.\@ 
Here, $\left( n_+ - n_- \right)$ is the spin polarization of the localized state, which depends on the ratio of $U_{0} / \Gamma$ for a single-orbital impurity. 
A generalization of the Anderson impurity model for a five-fold degenerate impurity state, as for $3d$ elements, predicts a transition from a magnetic to a non-magnetic impurity state for $(U_0 + 4 J) / \Gamma \le \pi$, where $J$ is the intra-atomic $d-d$ exchange \cite{Yosida65}.\@
\newline
For a quantitative analysis of $(U_0 + 4 J) / \Gamma$, we calculated $\tilde{U} = \left( U_{0} - J \right) \left( n_+ - n_-\right)$ in a self-consistent scheme for the multi-orbital systems $\mathrm{CrAu}_{n}^+$ \cite{Cococcioni05,Kulik10}.\@ Here, $\tilde{U}$ is derived from the slope of the linear response of the occupation number of the $3d$ impurity states to a rigid potential shift introduced at the impurity site \cite{Cococcioni05}.\@ 
Since $\tilde{U}$ is obtained from \textit{ab initio} calculations, the Coulomb interaction among the impurity $3d$ electrons, and all screening and hybridization effects are intrinsically accounted for. 
Hence, $U_0$ can be determined from $U_0 = \tilde{U} / \left( n_+ - n_- \right) + J$.\@ Since the atomic $d-d$ exchange interaction $J$ is only weakly screened, $J$ can be assumed to be independent of the cluster size and is of the order of $\unit[0.5-1]{eV}$ \cite{Cowan81,Sasioglu11}.\@ The half width $\Gamma$ of the localized impurity states is estimated as the weighted standard deviation of the $d$-projected density of states of the impurity.\@ The resulting values of $(U_0 + 4 J) / \Gamma$ obtained with constant $J = \unit[\left( 0.75 \pm 0.25 \right)]{eV}$ are shown in Fig.\ \ref{fig:onsiteU} (c).\@ As can be seen, the Anderson criterion for a magnetic impurity state is well satisfied throughout the whole size range, which is in perfect agreement with the non-vanishing XMCD signal presented in Fig.\ \ref{fig:CrAu_XMCD} for all $\mathrm{CrAu}_n^+$ clusters studied here. In particular, local maxima of $(U_0 + 4 J) / \Gamma$ are found for $\mathrm{CrAu}_{2}^+$ and $\mathrm{CrAu}_{6}^+$, where the width $2 \Gamma$ is reduced because of the large energy gap $\Delta E$ in the gold host.
\newline
The impurity spin magnetic moment that is predicted within the Anderson impurity model for these parameters $U_0$, $J$, and $\Gamma$ can be obtained by solving 
\begin{equation}
n_\pm=\frac{1}{\pi} \arctan\left[\frac{U_0+4J}{\Gamma}\left( n_\mp-0.5\right)\right]+0.5
\label{eq:AIM}
\end{equation}
\cite{Anderson61,Yosida65} for the values of $(U_0+4J) / \Gamma$ given in Fig.\ \ref{fig:onsiteU} (c).\@ As can be seen in Fig.\ \ref{fig:onsiteU} (a), the size dependence of the spin magnetic moment obtained from a DFT population analysis shows qualitative agreement with the spin magnetic moment deduced within the Anderson impurity model, and both are in accordance with the experimental data. Hence, the magnetic moments observed in $\mathrm{CrAu}_n^+$ clusters can essentially be explained in terms of the Anderson impurity model. Interestingly, however, it seems that the Anderson impurity model tends to underestimate the impurity magnetic moment. Beyond the simplified picture given here, a fully quantitative description might be obtained by an extension of the Anderson impurity model to explicitly include a discrete density of states that is absent in the bulk free-electron gas. 
\newline
Fig.\ \ref{fig:onsiteU} shows that the variation of the impurity magnetic moment is only of the order of 10~\% for $\mathrm{CrAu}_{n}^{+}$ even though the energy gap $\Delta E$ and the interaction $E_{\text{int}}$ vary by a factor of two. This is because even in bulk gold, i.e., in the absence of an energy gap, the chromium impurity is magnetic \cite{Brewer04} and carries a spin magnetic moment of $\unit[3.61]{\mu_B}$ \cite{FrotaPessoa04}.\@ The effect of reduced hybridization on this already large magnetic moment is limited, the more so as relation (\ref{eq:AIM}) does not translate the energy gap linearly to the spin polarization \cite{Anderson61,Yosida65}.\@ A much larger effect of the energy gap or a discrete density of states on the local magnetic moment should be observed in systems that are closer to or even below the threshold of quenching in the bulk case. In this case, the energy gap might even serve to restore the local magnetic moment of the impurity atom. 
\newline
In summary, the experimentally observed size dependence of the XMCD spectra of size-selected chromium-doped gold clusters are in line with the Anderson impurity model. The size-dependent variation of the spin magnetic moment can be linked to the amount of hybridization of the impurity with the host density of states and is governed by the energy gap of the host gold cluster. Electronic shell closure in the gold host leads to wide energy gaps $\Delta E$ in the free-electron states, which reduces the interaction with the impurity and causes the maximum spin magnetic moments of $\unit[5]{\mu_B}$ for $\mathrm{CrAu_{2,6}^+}$.\@ This effect is a result of quantum confinement and is unique to finite systems.
\newline
Beamtime for this project was granted at BESSY II beamline UE52-SGM, operated by Helmholtz-Zentrum Berlin. Technical assistance and user support by P.~Hoffmann and E.~Suljoti is gratefully acknowledged. Calculations were carried out on the DFG FOR 1282 computing cluster. The superconducting solenoid was provided by the Special Cluster Research Project of Genesis Research Institute, Inc.\@ We thank L.~Leppert for fruitful discussions. BvI acknowledges travel support by HZB.\@

\end{document}